%% file: main.tex
\documentclass{lipics-v2019}
\usepackage{thm-restate}
\hideLIPIcs  

\author{Supratik Chakraborty}{Department of Computer Science and Engineering\and IIT Bombay, India}{supratik@cse.iitb.ac.in}{}{}

\author{S. Akshay}{Department of Computer Science and Engineering\and  IIT Bombay, India}{akshayss@cse.iitb.ac.in}{}{}

\authorrunning{S. Chakraborty and S. Akshay} 

\Copyright{S. Chakraborty and S. Akshay} 



\relatedversion{Full version of paper accepted at MFCS'2022.} 

\nolinenumbers 

\title{On synthesizing computable Skolem functions for first order logic}

\input{macro}

\begin{document}
\maketitle

\begin{abstract}
Skolem functions play a central role in the study of first order logic, both from theoretical and practical perspectives. While every Skolemized formula in first-order logic makes use of Skolem constants and/or functions, not all such Skolem constants and/or functions admit effectively computable interpretations. Indeed, the question of whether there exists an effectively computable interpretation of a Skolem function, and if so, how to automatically synthesize it, is fundamental to their use in several applications, such as planning, strategy synthesis, program synthesis etc. 

In this paper, we investigate the computability of Skolem functions and their automated synthesis in the full generality of first order logic. We first show a strong negative result, that even under mild assumptions on the vocabulary, it is impossible to obtain computable interpretations of Skolem functions. We then show a positive result, providing a precise characterization of first-order theories that admit effective interpretations of Skolem functions, and also present algorithms to automatically synthesize such interpretations. We discuss applications of our characterization as well as complexity bounds for Skolem functions (interpreted as Turing machines).
\end{abstract}

\section{Introduction}\label{sec:intro}
\input{intro}

\section{Preliminaries}\label{sec:prelim}
\input{prelim}

\section{An illustrative example}\label{sec:example}
\input{example}

\section{Problem statement}\label{sec:problem}
\input{problem}

\section{Hardness of $\existsprob$ and $\synthprob$}
\label{sec:generalprob}
\input{existsprob}

\medskip

\noindent{\bf Fixing the formula.}
\input{undec}

\input{fixed-structure}
\input{complexity}

\section{Conclusion}\label{sec:conclusion}
\input{conclusion}

\bibliography{ref}

\end{document}

%% file: macro.tex
\usepackage{amsmath}
\usepackage{amssymb}
\usepackage{tikz}
\usetikzlibrary{shapes,arrows,fit,calc,positioning,automata}

\newcommand{\nonl}{\renewcommand{\nl}{\let\nl\oldnl}}

\makeatletter
\newif\if@restonecol
\makeatother

\usepackage[ruled,vlined,linesnumbered,lined]{algorithm2e}
\usepackage{algorithmic}

\usepackage{xcolor,colortbl}
\definecolor{Gray}{gray}{0.95}
\definecolor{LightCyan}{rgb}{0.88,1,1}

 \usepackage{relsize}

\newcommand{\X}{\ensuremath{\mathbf{X}}}
\newcommand{\Y}{\ensuremath{\mathbf{Y}}}
\newcommand{\Z}{\ensuremath{\mathbf{Z}}}
\newcommand{\U}{\ensuremath{\mathbf{U}}}

\newcommand{\bF}{\ensuremath{\mathbf{F}}}

\newcommand{\MM}{\ensuremath{\mathfrak{M}}}
\newcommand{\univ}{\ensuremath{\mathcal{U}}}

\newcommand{\ttrue}{\ensuremath{\mathsf{true}}}
\newcommand{\ffalse}{\ensuremath{\mathsf{false}}}

\newcommand{\eat}[1]{}

\newcommand{\Ff}{\mathcal F}

\newcommand{\subst}[3]{\ensuremath{{#1}[{#2}\mapsto {#3}]}}  
\newcommand{\modelconst}[1]{\ensuremath{{#1}_{C}}}
\newcommand{\eldiag}[1]{\ensuremath{\mathcal{ED}({#1})}}
\newcommand{\diag}[1]{\ensuremath{\mathcal{D}({#1})}}
\newcommand{\expvoc}[2]{\ensuremath{{#1}({#2})}}
\newcommand{\Theory}[1]{\ensuremath{\mathit{Th}({#1})}}

\newcommand{\existsprob}{{ \textsc{SkolemExist}}}
\newcommand{\synthprob}{{ \textsc{SkolemSynthesis}}}

\renewcommand{\implies}{\Rightarrow}

\newcommand{\mcomment}[1]{}

\newcommand{\VV}{\mathcal{V}}

\newcommand{\Univ}{U}

\usetikzlibrary{decorations.pathreplacing} 
\usetikzlibrary{patterns}

\usepackage[ruled,vlined,linesnumbered,lined]{algorithm2e}

\tikzset{
block/.style={
  draw, 
  rectangle, 
  minimum height=0cm, 
  minimum width=2cm, align=center,
  }, 
line/.style={->,>=latex'}
}
\tikzstyle{galivenode}=[circle,fill=green!50!black,thick,inner sep=1pt,minimum size=4mm]
\tikzstyle{ralivenode}=[circle,fill=red!70!black,thick,inner sep=1pt,minimum size=4mm]

\tikzstyle{alivenode}=[circle,fill=black!80,thick,inner sep=1pt,minimum size=4mm]
\tikzstyle{deadnode}=[circle,fill=black!10,thick,inner sep=0pt,minimum size=4mm]
\tikzstyle{rdnode}=[circle,draw,fill=red!10,thick,inner sep=2pt,minimum size=4mm]
\tikzstyle{grnode}=[circle,draw,fill=green!10,thick,inner sep=2pt,minimum size=4mm]

\tikzstyle{ynode}=[rectangle,draw=black,fill=yellow!10,thick,inner sep=3pt,minimum size=4mm]
\tikzstyle{gnode}=[rectangle,fill=green!10,thick,inner sep=1pt,minimum size=4mm]
\tikzstyle{bnode}=[rectangle,fill=red!10,thick,inner sep=1pt,minimum size=4mm]

\newcommand{\Algo}[1]{\ensuremath{\mathcal{A}^{[#1]}}}
\newcommand{\Alg}{\Algo{F}}
\newcommand{\AlgP}{\Algo{P}}

\newcommand{\TM}[1]{\ensuremath{\mathsf{TM}_{{#1}}}}

%% file: intro.tex
The history of Skolem functions can be traced back to 1920, when the Norwegian
mathematician, Thoralf Albert Skolem, gave a simplified proof of a
landmark result in logic, now known as the \emph{L\"{o}wenheim-Skolem}
theorem.   Skolem's proof made
use of a key observation: \emph{For every first order logic formula
$\exists y\,\varphi(x, y)$, the choice of $y$ that makes $\varphi(x,
y)$ true (if at all) depends on $x$ in general. This dependence can be
thought of as implicitly defining a function that gives the ``correct''
value of $y$ for every value of $x$.  If $F_y$ denotes a fresh unary function
symbol, the second order sentence 
$\exists F_y\,\forall x\, \big(\exists y\,\varphi(x,
y) \Rightarrow \varphi(x, F_y(x))\big)$ formalizes this idea.}  Since the
implication trivially holds in the other direction too, the second order sentence  $\exists F_y\,\forall x\, \big(\exists y\,\varphi(x,
y) \Leftrightarrow \varphi(x, F_y(x))\big)$ is valid.

Let $\xi_1 \equiv \exists  y\, \varphi(x, y)$ and $\xi_2 \equiv \varphi(x, F_y(x))$. The fresh function symbol $F_y$ introduced in transforming $\xi_1$ to $\xi_2$ is called a \emph{Skolem function}.  Skolem functions play an extremely important role in logic -- both from theoretical and applied perspectives.  
While it suffices in some contexts to simply know that a Skolem function $F_y$ exists, in other contexts, we require an effective procedure to compute $F_y(x)$ for every value of $x$.   This motivates us to ask if Skolem functions are always computable, and whenever they are, can we  algorithmically generate a halting Turing machine that computes the function? Note that we are concerned with computability at two levels here: (i) computability of the Skolem function itself, and (ii) computability of a halting Turing machine that computes the Skolem function.  For clarity of exposition, we call the Turing machine referred to in (ii) above a \emph{computable interpretation of Skolem function}, and the problem of generating it algorithmically the \emph{synthesis problem for Skolem functions}.

The synthesis problem for Skolem functions has been studied in detail in the propositional setting, specifically for quantified Boolean formulas (QBF) with a $\forall^*\exists^*$ quantifier prefix~\cite{Jian,bierre,FHSB17,fmcad2015:skolem,rsynth,cadet,rsynth:fmcad2017,tacas2017,RabeTRS18,cav2018,fmcad19,Rabe19,fmsd20,GRM20}.  Surprisingly, a similar in-depth investigation in the context of general first order logic appears lacking in the literature, despite several potential applications, viz. automatic program synthesis and repair~\cite{SG10,pldi10,VR11}. Some notable works in the context of specific theories include those of Kuncak et al~\cite{pldi10} (for linear rational arithmetic), Spielman et al~\cite{SK12} (unbounded bit-vector theory), Preiner et al~\cite{PNB17} (bit-vector theory) etc. in which terms that serve as interpretations of Skolem functions in specific theories are synthesized. In~\cite{Jian}, Jiang presented a partial approach for quantifier elimination in general first-order theories by relying on the availability of functions that can be conditionally expressed by a finite set of terms. Unfortunately, such finite conditional decomposition may not always be possible (as acknowledged in~\cite{Jian}), even when a computable interpretation of the Skolem function exists.  The problem of quantifier-free constraint solving, i.e. finding assignments of free variables that render a quantifier-free formula $\ttrue$, has been investigated in depth for several theories, viz. propositional logic, theory of arrays, linear rational arithmetic, real algebraic numbers, Presburger arithmetic, regular languages of finite strings etc. If the theory also admits effective quantifier elimination, this yields an algorithm for synthesizing computable interpretations of Skolem functions.  However, not all first order theories admit effective quantifier elimination, e.g. Presburger arithmetic (without divisibility predicates) or the theory of evaluated trees~\cite{DK06} does not. We show that even for such theories, computable interpretations of Skolem functions can be synthesized algorithmically.  

Our main contributions are to ask and answer the following questions:
\begin{itemize}
    \item Does there always exist computable interpretations of Skolem functions for a first order formula interpreted over a structure? We answer this question strongly in the negative by showing that uncomputable interpretations cannot be avoided even with one binary predicate and one existential quantifier in the formula.
    \item We next ask if it is possible to algorithmically decide whether computable interpretations exist for all Skolem functions, given a formula and a structure over which it is interpreted.  We answer this question in the negative.
    \item Next, we ask if it is possible to characterize the class of structures such that effectively computable interpretations of Skolem functions can be algorithmically synthesized for all formulas interpreted over a structure in the class.  We answer this by showing that decidability of the elementary diagram of a structure serves as the required necessary and sufficient condition. Using this result, we show that several important first-order theories admit synthesis of effectively computable Skolem functions, while others do not.
    \item For structures satisfying the condition in the above characterization, we present lower and upper complexity bounds for effectively computable interpretations of Skolem functions.  
    \item  Finally, we distinguish between synthesizing Skolem functions as halting Turing machines vs terms in the underlying logic and show that the latter is a strictly weaker notion.
\end{itemize} 
Our results reveal a highly a nuanced picture of the computability landscape for synthesizing interpretations of Skolem functions in first-order logic. We hope that this work will be a starting point towards further research into the design of practical algorithms (whenever possible) to synthesize Skolem functions for various first order theories.

%% file: prelim.tex
As Turing machine with tape alphabet $\{0,1\}$ can be encoded as a natural number (we use $\mathbb{N}$ for naturals), and every finite string over $\{0,1\}^*$ can be encoded as a natural number, we often speak of Turing machine $i$, denoted $\TM{i}$, running on input string $j$, where $i, j \in \mathbb{N}$.

We use  $x$, $y$, $z$, etc., possibly with subscripts, to denote first order variables, $\X$, $\Y$, $\Z$, etc., possibly with subscripts, to denote sequences of first order variables.  $\varphi$, $\xi$, $\alpha$, possibly with subscripts, are used to denote formulas.  For a sequence $\X_i$, $|\X_i|$ denotes the count of variables in $\X_i$, and $x_{i,1}, \ldots x_{i,|\X_i|}$ denotes the variables. A \emph{vocabulary} $\VV$, is a set of function and/or predicate symbols, along with their respective arities. Constants are function symbols with arity $0$. We  assume that \emph{$\VV$ has finitely many predicate and function symbols, except possibly for countably infinitely many constant symbols.}  We also assume that \emph{a special binary predicate "$=$" (equality) is present in every vocabulary}.

We consider first order logic formulas over vocabulary $\VV$, also called \emph{$\VV$-formulas}. The notion of \emph{bound} and \emph{free} variables is standard, $\VV$-formulas without free variables are \emph{$\VV$-sentences}.  A \emph{$\VV$-term} is either a variable or $f(t_1, \ldots t_k)$, where $f$ is a $k$-ary function symbol in $\VV$ and $t_1, \ldots t_k$ are $\VV$-terms.   When $\VV$ is implicit from the context, we omit it.  A \emph{ground term} (resp. \emph{ground formula}) is a term (resp. formula) without any variables. For $x$, a free variable in $\xi$, $t$ a term in which all variables (if any) are free in $\xi$, $\subst{\xi}{x}{t}$ denotes the formula obtained by \emph{substituting $t$ for $x$ in $\xi$}, i.e., replacing every free occurrence of $x$ in $\xi$ with $t$. 

A \emph{$\VV$-structure} $\MM$ consists of a \emph{universe} $U^{\MM}$ of elements and an interpretation of every predicate and function symbol in $\VV$ over $U^{\MM}$.  The interpretation of the special predicate "=" is always  the identity relation, and we write $t_1 = t_2$ instead of $=(t_1,t_2)$ for notational convenience.  We denote the interpretation of a predicate symbol $P$ (resp. function symbol $f$) in $\MM$ as $P^{\MM}$ (resp. $f^{\MM}$).  In general, an interpretation of a predicate or function symbol may be well-defined but not computable. We say a $\VV$-structure $\MM$ is \emph{computable} if $U^{\MM}$ is countable and if $P^{\MM}$ (resp. $f^{\MM}$) is computable for all predicate symbol $P$ (resp. function symbol $f$) in $\VV$.  In other words, there exists a halting Turing machine for computing the interpretations $P^{\MM}$ (resp. $f^{\MM}$). \emph{Throughout this paper, we assume that all $\VV$-structures are computable.} This is motivated by practical applications of Skolem functions; additionally, non-computable $\VV$-structures may make it difficult (even impossible) to obtain computable interpretations of Skolem functions in most cases. A computable $\VV$ structure can be finitely represented, e.g. by using a single bit to encode whether the universe is finite or countably infinite, and by giving a $\mathbb{N}$-encoding of each Turing machine that computes an interpretation of a predicate or function symbol. If  there are countably infinite constant symbols, we assume that interpretations of all of them can be collectively encoded by a single Turing machine that computes a mapping from $\mathbb{N}$ (index of constant symbol) to $\mathbb{N}$ (index of element in universe).  If a $\VV$-formula $\xi(\Z)$ evaluates to $\ttrue$ when interpreted over $\MM$ and with $\Z$ set to $\sigma \in \big(U^{\MM})^{|\Z|}$, we say that $\MM$ is a \emph{model} of $\xi(\sigma)$ and denote it by $\MM \models \xi(\sigma)$. An \emph{expansion} of a vocabulary $\VV$ is a vocabulary $\VV'$ such that $\VV \subseteq \VV'$.  Given a $\VV$-structure $\MM$ and a $\VV'$-structure $\MM'$, where $\VV'$ is an expansion of $\VV$, $\MM'$ is an \emph{expansion} of $\MM$ if (i) $U^{\MM'} = U^{\MM}$, and (ii) all predicate/function symbols in $\VV$ are interpreted identically in $\MM$ and $\MM'$.

For a quantifier $Q\in\{\exists, \forall\}$ and sequence of variables $\X_i = (x_{i,1}, \ldots x_{i,|\X_i|})$, we use $Q \X_i$ to denote the block of quantifiers $Q x_{i,1}\, \ldots Q x_{i,|\X_i|}$. Every first order logic formula can be effectively transformed to a semantically equivalent \emph{prenex normal form}, in which all quantifiers appear to the left of the quantifier-free part of the formula.
Henceforth, we assume all first order formulas are in prenex normal form, unless stated otherwise. Let $\xi(\Z) \equiv \forall \X_1 \exists \Y_1 \cdots \forall \X_q \exists \Y_q\, \varphi(\Z, \X_1, \Y_1, \ldots \X_q, \Y_q)$ be such a formula, where $\Z$ is a sequence of free variables,  and $\varphi$ is  quantifier-free. We say that $\forall \X_1 \exists \Y_1 \cdots \forall \X_q \exists \Y_q$ is the \emph{quantifier prefix} of the formula, and it has $q$ $\forall^* \exists^*$ blocks.  The quantifier-free part, i.e. $\varphi$, is called the \emph{matrix} of the formula.  Note that in case the leading (leftmost) quantifier in $\xi$ is existential, $\X_1$ may be considered to be an empty sequence, and similarly, if the trailing (rightmost) quantifier in $\xi$ is universal.  Every variable $y_{i,j}$ that is existentially quantified in the quantifier prefix is called an an \emph{existential variable} in $\xi$.  The notion of \emph{universal variables} is analogously defined.  The quantifier prefix imposes a total order on the quantified variables in $\xi$.  We say that a variable $u$ is \emph{to the left} (resp. \emph{right}) of variable $v$ in the quantifier prefix iff $Qu$ appears to the left (resp. right) of $Q'v$ in the quantifier prefix, where $Q, Q' \in \{\exists, \forall\}$. 
\medskip 

\noindent{\bf Skolemization.}
Given a formula $\xi$ in prenex normal form, \emph{Skolemization} refers to the process of transforming $\xi$ to a new formula $\xi^\star$ via the following steps: (i) for every existential variable $y_{i,j}$, substitute $F_{y_{i,j}}(\Z, \X_1, \ldots \X_i)$ for $y_{i,j}$ in $\varphi$, where $F_{y_{i,j}}$ is a new function symbol of arity $|\Z| + \sum_{j=1}^{i}|\X_j|$, and (ii) remove all existential quantifiers from the quantifier prefix of $\xi$. The functions $F_{y_{i,j}}$ introduced above are called \emph{Skolem functions}.  In case $\xi$ has no free variables and the leading quantifier is existential, the Skolem functions for variables in the leftmost existential quantifier block have no arguments (i.e. they are nullary functions).  Such functions are also called \emph{Skolem constants}.  The sentence $\xi^\star$ is said to be in \emph{Skolem normal form} if the matrix of $\xi^\star$ is in conjunctive normal form.  The key guarantee of Skolemization is as follows: for every existential variable $y_{i,j}$, let $\xi^\star_{y_{i,j}}$ denote the formula obtained by Skolemizing all existential variables to the left of $y_{i,j}$ in the quantifier prefix.  Formally, $\xi^\star_{y_{i,j}}$ is obtained by (i) substituting the Skolem function $F_{y_{k,l}}$ for every existential variable $y_{k,l}$ to the left of $y_{i,j}$ in the quantifier prefix, and (ii) removing all quantifiers to the left of and including $\exists y_{i,j}$ from the quantifier prefix. Note that $\xi^\star_{y_{i,j}}$ has free variables in $\Z, \X_1, \ldots \X_i, y_{i,j}$. Skolemization guarantees that for every $\VV$-structure $\MM$ over which $\xi$ is interpreted, there always exists an expansion $\MM^\star$ of $\MM$ that provides an interpretation of $F_{y_{i,j}}$ for all existential variables $y_{i,j}$ such that the following holds for every $i \in \{1, \ldots q\}$ and $j \in \{1, \ldots |\Y_i|\}$: 
\begin{align}
\forall \Z \forall \X_1 \ldots \forall \X_i\, \big(\exists y_{i,j}\, \xi^\star_{y_{i,j}} ~\Leftrightarrow~ \subst{\xi^\star_{y_{i,j}}}{y_{i,j}}{F_{y_{i,j}}(\Z, \X_1, \ldots \X_i)}\big) \label{fact1}
\end{align} 

\begin{example}\label{example:skolem}
Consider $\xi(z) \equiv \exists y \forall x \exists u \forall v \exists w\, \varphi(z, x, y, u, v, w)$.   Skolemizing gives
$\xi^\star \equiv \forall x \forall v\, \varphi(z, x, F_{y}(z), F_{u}(z, x), v, F_{w}(z, x, v))$, where $F_{y}(z), F_{u}(z, x)$ and $F_{w}(z, x, v)$ are Skolem functions for $y$, $u$ and $w$ respectively.  Using the notation introduced above, we have
\begin{itemize}
 \item $\xi^\star_y(z, y) ~\equiv~ \forall x \exists u \forall v \exists w\, \varphi(z, x, y, u, v, w)$
\item $\xi^\star_u(z, x, u) ~\equiv~ \forall v \exists w\, \varphi(z, x, F_y(z), u, v, w)$
\item $\xi^\star_w(z, x, v, w) ~\equiv~ \varphi(z, x, F_y(z), F_u(z, x), v, w)$
\end{itemize}
By virtue of Skolemization, we know that for every structure $\MM$ over which $\xi$ is interpreted, there
exists an expansion $\MM^\star$ that interprets $F_y$, $F_u$ and $F_w$ such that the following hold.
\begin{itemize}
    \item $\forall z \big(\exists y\, \xi^\star_y(z, y) \Leftrightarrow \subst{\xi^\star_y}{y}{F_y(z)}\big)$
    \item $\forall z \forall x\, \big(\exists u\, \xi^\star_u(z, x, u) \Leftrightarrow  \subst{\xi^\star_u}{u}{F_u(z, x)}\big)$
    \item $\forall z \forall x \forall v\, \big(\exists w \xi^\star_w(z, x, v, w) \Leftrightarrow \subst{\xi^\star_w}{w}{F_w(z, x, v)}\big)$ 
\end{itemize}
\end{example}

Let $\VV^\star$ be the expansion of $\VV$ obtained by adding all Skolem function and constant symbols in $\xi^\star$ to $\VV$.
In general, a $\VV$-structure $\MM$ over which $\xi(\Z)$ is interpreted can be expanded to a $\VV^\star$-structure by adding interpretations of Skolem functions for all existential variables in $\xi(\Z)$.  However, not every such expansion of $\MM$ may model the sentence (\ref{fact1}) above for every existential variable $y_{i,j}$.  Skolemization guarantees that there exists at least one "correct" expansion $\MM^\star$ of $\MM$ that does so. We call the interpretation of Skolem functions in such a "correct" expansion as an \emph{$\MM$-interpretation} of the Skolem functions.  There may be multiple "correct" expansions of $\MM$, and hence multiple $\MM$-interpretations of Skolem functions. Skolemization guarantees the existence of at least one $\MM$-interpretation of all Skolem functions/constants; however, it doesn't tell us whether these are computable interpretations, and if so, can we algorithmically synthesize the interpretation as a halting Turing machine? These are two central questions that concern us in this paper.

Sometimes, given a $\VV$-formulas $\xi$,  we can find an $\MM$-interpretation of Skolem functions that works in the same way for all computable structures $\MM$ over which $\xi$ is interpreted (modulo differences in interpreting predicates and functions).  Formally, suppose there exists a halting Turing machine $M_\xi$ with access to oracles that compute the interpretations of predicates and functions in $\MM$, and suppose $M_\xi$ computes an $\MM$-interpretation of Skolem functions for all existential variables in $\xi$, and for all $\VV$-structures $\MM$. Then, we say that $\xi$ admits a \emph{uniform representation} of $\MM$-interpretations of Skolem functions.
\medskip

\noindent{\bf Model theory.}\label{sec:modeltheory}
We use $\expvoc{\VV}{\MM}$ to denote the expansion of $\VV$ obtained by adding a fresh constant symbol $c_e$ for every element $e \in U^{\MM}$, if not already present in $\VV$.  Clearly, if $U^{\MM}$ and $\VV$ are countable, so is $\expvoc{\VV}{\MM}$.  We use $\modelconst{\MM}$ to denote the expansion of $\MM$ to a $\expvoc{\VV}{\MM}$-structure that interprets the additional constants in $\expvoc{\VV}{\MM}$ in the natural way, i.e. $c_e$ is interpreted to have the value $e$, for all $e \in U^{\MM}$.  The \emph{elementary diagram} of $\MM$, denoted $\eldiag{\MM}$, is the set of all $\expvoc{\VV}{\MM}$-sentences $\xi$ such that $\modelconst{\MM} \models \xi$.  The \emph{diagram} of $\MM$, denoted $\diag{\MM}$, is the set of all literals in $\eldiag{\MM}$, i.e. the set of all atomic ground formulas that hold in $\modelconst{\MM}$.  Clearly, $\diag{\MM} \subseteq \eldiag{\MM}$.
A set $\Gamma$ of $\VV$-sentences is called a \emph{$\VV$-theory} if it is consistent, i.e. there exists a $\VV$-structure that serves as a model for every sentence in $\Gamma$.  Given a $\VV$-structure $\MM$, the set of all first order $\VV$-sentences $\xi$ such that $\MM \models \xi$ is called the \emph{theory of $\MM$}, denoted $\Theory{\MM}$.  Note that both $\eldiag{\MM}$ and $\diag{\MM}$ are $\VV$-theories, and $\eldiag{\MM} = \Theory{\modelconst{\MM}}$, where $\Theory{\modelconst{\MM}}$ is the $\expvoc{\VV}{\MM}$-theory of $\modelconst{\MM}$. 
We say that a $\VV$-theory $\Gamma$ is \emph{decidable} iff there exists a  Turing machine that takes as input an arbitrary $\VV$-sentence $\xi$ and always halts and correctly reports whether $\xi \in \Gamma$ or not.  If $\MM$ is a computable structure, it follows immediately that $\diag{\MM}$ is a decidable theory, but $\eldiag{\MM}$ is not necessarily so.

A $\VV$-theory $\Gamma$ is said to admit \emph{quantifier elimination} if for every $\VV$-formula $\xi(\Z)$ with free variables $\Z$, there exists a semantically equivalent quantifier-free  $\VV$-formula $\xi^{\#}(\Z)$ such that the sentence $\forall \Z\, \big(\xi(\Z) \Leftrightarrow \xi^{\#}(\Z)\big)$ is in $\Gamma$.  If, in addition, there exists a Turing machine that takes an arbitrary $\VV$-formula ($\xi$) as input and computes its quantifier-eliminated form ($\xi^{\#}$) and halts, we say that $\Gamma$ admits \emph{effective quantifier elimination}\footnote{There is a technique, popularly called "Morleyization", that trivially makes a theory admit effective quantifier elimination by expanding the vocabulary to include a separate predicate symbol for each $\VV$-formula.  For purposes of this paper, we disallow expansion of the vocabulary (and hence "Morleyization") during effective quantifier elimination.}. For a $\VV$-structure $\MM$, we say that $\Theory{\MM}$ admits \emph{effective constraint solving} if there exists a Turing machine that takes a $\VV$-formula $\xi(\Z)$ with free variables $\Z$ as input and halts after reporting one of two things: (i) a $|\Z|$-tuple $\sigma$ of elements from $U^{\MM}$ such that $\MM \models \xi(\sigma)$, or (ii) no such $|\Z|$-tuple of elements from $U^{\MM}$ exists.  Note that the formula $\xi(\Z)$ may have quantifiers in general.  In case the above Turing machine exists only if $\xi(\Z)$ is  quantifier-free, we say that $\Theory{\MM}$ admits \emph{effective quantifier-free constraint solving}.  
Clearly, if $\Theory{M}$ admits effective quantifier elimination and effective quantifier-free constraint solving, then it also admits effective constraint solving.

%% file: example.tex
Consider the vocabulary $\VV = \{P, c, d\}$, where $P$ is a binary predicate symbol, and $c$ and $d$ are constants, and the first-order $\VV$-sentence $\xi \equiv \forall x \exists y P(x, y) \wedge \big(P(x, c)\vee P(x, d)\big)$. We will use $\varphi(x,y)$ to denote the matrix of the above formula, i.e. $P(x, y) \wedge \big(P(x, c)\vee P(x, d)\big)$.  On Skolemizing $\xi$ we get $\xi^\star \equiv \forall x \varphi(x, F_y(x))$, where $F_y$ is a fresh unary Skolem function symbol.  Let $\MM$ be a computable $\VV$-structure. We now ask if there exists an algorithm $\Alg$ that serves as a computable interpretation of $F_y: U^{\MM} \rightarrow U^{\MM}$.   A careful examination of $\xi$ and $\xi^\star$ reveals that such an algorithm indeed exists. Specifically, the algorithm (represented informally as an imperative "program" for ease of understanding) "{\bfseries input($x$); if $\mathtt{P^{\MM}(x, c^{\MM})}$ then return $c^{\MM}$ else return $d^{\MM}$}" takes as input $x \in U^{\MM}$ and returns either $c^{\MM}$ or $d^{\MM}$ depending on whether $P^{\MM}(x, c^{\MM})$ evaluates to $\ttrue$ or $\ffalse$.  If we let this algorithm interpret $F_y$ in the expansion $\MM^\star$ of $\MM$, then it is not hard to see that we indeed have $\MM^\star \models \forall x \big(\exists y \varphi(x,y) \Leftrightarrow \varphi(x, F_y(x))$.

However, is this always possible? Consider the $\VV$-formula $\alpha \equiv \forall x \exists y\, P(x, y)$ instead of $\xi$, whose Skolemized version is $\alpha^\star \equiv \forall x\, P(x, F_y(x))$. As we show in Section~\ref{sec:generalprob}, it is impossible to obtain a computable $\MM$-interpretation of the Skolem function $F_y(x)$ in this case for all $\VV$-structures $\MM$. 

There are several observations that one can now make. Clearly, algorithm $\Alg$ described above is specific to the formula $\xi$; a different formula would have required a different algorithm to be designed for its Skolem function(s).  Interestingly, algorithm $\Alg$ also requires access to the interpretations of $c$, $d$ and $P$ in the $\VV$-structure $\MM$ on which $\xi$ is interpreted.  Since we are given an effectively computable interpretation of $P$ in $\MM$, there exists an algorithm $\AlgP$ to compute $P^{\MM}$.  Algorithm $\Alg$ effectively uses $\AlgP$ as a sub-routine to compute the value of $F_y(x)$ for every $x \in U^{\MM}$.  Note that if the interpretation of $P$ (in perhaps a different $\VV$-structure $\MM'$) was not effectively computable, the "program" above would not serve as an effectively computable interpretation of $F_y$.  This underlines the importance of effectively computable structures in the synthesis of Skolem functions.  

It is easy to see that "{\bfseries input(x); if $\mathtt{P^{\MM}(x, c^{\MM})}$ then return $c^{\MM}$ else return $d^{\MM}$}" \emph{uniformly} serves as a computable interpretation of $F_y$ in every computable $\VV$-structure $\MM$ over which $\xi$ is interpreted.  Regardless of the actual structure $\MM$, a computable $\MM$-interpretation of $F_y$ is obtained by invoking algorithms to compute interpretations of $P$, $c$, $d$ in $\MM$ as sub-routines.  Thus we get a uniform representation of an $\MM$-interpretation of $F_y$.  

Finally, the interpretation of Skolem function $F$ discussed above is represented as an algorithm, and not as a $\VV$-term.  Is it possible to obtain a $\VV$-term that uniformly represents an $\MM$-interpretation of $F_y$ in this case? To answer this, first observe that there are only two terms, viz. $c$ and $d$, that can be formed using $\VV$. If one of these terms serves as a uniform $\MM$-interpretation of $F_y$, choose a structure $\MM$ as follows: $U^{\MM} = \{a_0, a_1\}, c^{\MM} = a_0, d^{\MM} = a_1, P^{\MM}(a_0, a_0) = P^{\MM}(a_1, a_1) = \ffalse$ and $P^{\MM}(a_0, a_1) = P^{\MM}(a_1, a_0) = \ttrue$. Clearly $\MM \models \forall x \exists y \varphi(x, y)$.  However, with $F_y(x) = c$ (or $F_y(x) = d$), we have $MM^\star \not\models \forall x \big(\exists y \varphi(x, y) ~\Leftrightarrow~ \varphi(x, F(x))\big)$.  Thus even when an effectively computable interpretation of a Skolem function exists, it may not be representable as a term over $\VV$.

%% file: problem.tex
We now formulate the primary questions that we wish to address in this paper. 
\begin{enumerate}
    \item Given a vocabulary $\VV$, a $\VV$-formula $\xi(\Z)$ in prenex normal form and a computable $\VV$-structure $\MM$, the $\existsprob$ problem asks if there exists a \emph{computable} $\MM$-interpretation of Skolem functions for all existential variables in  $\xi$. We have already seen in Section~\ref{sec:example} that there are positive instances of $\existsprob$.  We ask if there are negative instances as well, i.e. there is no computable $\MM$-interpretation of Skolem functions.
    \item Next, we ask if  $\existsprob$ is decidable.  
    \item We then consider special cases where either the formula $\xi(\Z)$ or structure $\MM$ is fixed, and ask if it is possible to characterize the class of problems where the $\existsprob$ problem has a positive answer.
    \item In cases where the $\existsprob$ problem has a positive answer, we ask the following:
    \begin{enumerate}
        \item Does there exist an algorithm to synthesize computable $\MM$-intepretations of Skolem functions?  We call this problem $\synthprob$ and consider two variants of it,  where either (i) $\VV$ and $\xi(\Z)$ are fixed and $\MM$ is the input of $\synthprob$, or (ii) $\VV$ and $\MM$ is fixed and $\xi$ is the input of $\synthprob$.
        \item Is it possible to obtain finite uniform representations of $\MM$-interpretations of Skolem functions, and if so, can we obtain these as $\VV$-terms?
        \item In case $\existsprob$ has a positive answer, can we give bounds on the worst-case running time of computable $\MM$-interpretations of Skolem functions?
    \end{enumerate}
\end{enumerate}
Note that $\synthprob$ is not meaningful in cases where  $\existsprob$ has a negative answer. Hence, we don't try to answer $\synthprob$ in negative instances of $\existsprob$.  Moreover, all the above problems except the last one is trivial if the universe $U^{\MM}$ is finite. Therefore, we focus mostly on structures with countably infinite universe.

%% file: existsprob.tex
We have already seen a positive instance (i.e. problem instance with positive answer) of $\existsprob$  in Section~\ref{sec:example}. The following lemma shows that $\existsprob$ always has a positive answer if all Skolem functions are Skolem constants. In the following, we use $(\VV, \MM, \xi)$ to denote an instance of $\existsprob$, where $\VV$ is a vocabulary, $\MM$ is a computable $\VV$-structure and $\xi$ is a $\VV$-formula.
\begin{restatable}{lemma}{existsprobconst}
\label{lem:existsprob_const}
For every vocabulary $\VV$, every computable $\VV$-structure $\MM$ and every $\VV$-sentence $\exists \Y\, \varphi(\Y)$, where $\varphi$ is a quantifier-free $\VV$-formula with free variables in $\Y$, the instance $(\VV, \MM, \xi)$ of $\existsprob$ has a positive answer.
\end{restatable}

\begin{proof}
Let $\varphi(c_{\Y})$ denote the Skolemized form of $\xi$, where $c_{\Y}$
is a vector of constant symbols.  We show that there always exists a halting Turing machine, say $M$, that computes $\sigma \in \big(U^{\MM}\big)^{|\Y|}$ such that $\MM \models \big(\exists \Y\, \varphi(\Y) ~\Leftrightarrow~ \varphi(\sigma)\big)$.
The proof is by case analysis.
\begin{itemize}
    \item If $\MM \models \exists \Y\,\varphi(\Y)$, then there exists $\sigma \in \big(U^{\MM}\big)^{|\Y|}$ such that $\MM \models \varphi(\sigma)$.  Now, since $U^{\MM}$ is countable, so is $\big(U^{\MM}\big)^{|\Y|}$. Therefore, for every $i \in \mathbb{N}$, there exists a Turing machine that outputs the $i^{th}$ element in the enumeration of $\big(U^{\MM}\big)^{|\Y|}$ and halts. Hence, there exists a Turing machine that outputs the $|\Y|$-tuple $\sigma$ referred to above and halts, and
    $\MM \models \big(\exists \Y\, \varphi(\Y) ~\Leftrightarrow~ \varphi(\sigma)\big)$.
    \item If $\MM \not\models \exists \Y\, \varphi(\Y)$, then for every $\sigma \in \big(U^{\MM}\big)^{|\Y|}$, we have $\MM \not\models \varphi(\sigma)$.
    Hence, $\MM \models \big(\exists \Y\, \varphi(\Y) ~\Leftrightarrow~ \varphi(\sigma)\big)$.  Thus, a Turing machine that outputs the first element in the enumeration of $\big(U^{\MM}\big)^{|\Y|}$ serves our purpose.
\end{itemize}
It follows that in all cases, there exists a halting Turing machine that computes $\sigma \in \big(U^{\MM}\big)^{|\Y|}$ such that $\MM \models \big(\exists \Y\, \varphi(\Y)$ $~\Leftrightarrow~$ $\xi(\sigma)\big)$.  Hence, the $(\VV, \MM, \xi)$ instance of $\existsprob$ has a positive answer.
\end{proof}

\noindent However, there are negative instances of $\existsprob$, even with a restricted vocabulary. 
\begin{restatable}{theorem}{existsprobneg}
\label{thm:existsprob_neg}
There exists a negative instance of $\existsprob$ where the vocabulary has a single binary predicate.
\end{restatable}
\begin{proof}
Let $\VV = \{Q\}$, where $Q$ is a binary predicate symbol. Consider the $\VV$-sentence 
$ \xi \equiv \forall u \exists v\, Q(u, v)$ and the $\VV$-structure $\MM$, where $U^{\MM} = \mathbb{N}$ and $Q^{\MM}(u, v) = \ttrue$ iff the Turing machine corresponding to natural number $u$, say $\TM{u}$, halts on the empty tape within $v$ steps.  Notice that $\MM$ is a computable structure, since for every $(u, v) \in \mathbb{N}^2$, we can simulate $\TM{u}$ on the empty tape for up to $v$ steps and check if it halts.  If so, $Q^{\MM}(u, v) = \ttrue$, otherwise $Q^{\MM}(u, v) = \ffalse$.  

On Skolemizing $\xi$, we get $x^\star \equiv \forall u\, Q(u, F_v(u))$, where $F_v$ is the Skolem function for $v$.
Now, suppose there exists a computable $\MM$-interpretation of $F_v$.  In other words, there exists a halting Turing machine, say $M$, that takes as input  $u \in \mathbb{N}$ and computes an output $v \in \mathbb{N}$.  
We claim that $M$ can be used to solve the halting problem on the empty tape.  Specifically, to find if Turing machine $\TM{i}$ halts on the empty tape, we run $M$ with input $u$ set to $i$.  Let $j \in \mathbb{N}$ be the value computed by $M$ for this input.  Then, from the guarantee of Skolemization, we know that $\MM \models \big(\exists v\, Q(i, v) ~\Leftrightarrow Q(i,j)\big)$.  We can now simulate $\TM{i}$ on the empty tape for up to $j$ steps and check if it halts by then.  If it does, we know that $\TM{i}$ halts on the empty tape.  Otherwise, $Q^{\MM}(i, j) = \ffalse$ and $\exists v\, Q^{\MM}(i, v)$ must also be $\ffalse$.  In other words, $\TM{i}$ does not halt on the empty tape.

As the halting problem on empty tape is undecidable, it follows that Turing machine $M$ cannot exist, i.e., there is no computable $\MM$-interpretation of $F_v$. Hence, the instance $(\VV, \MM, \xi)$ of $\existsprob$ has a negative answer.
\end{proof}

Now that we know there are positive and negative instances of $\existsprob$, we ask if  $\existsprob$ is decidable. Unfortunately, we obtain a negative answer in general.
\begin{restatable}{theorem}{existsprobundec}
\label{thm:existsprob_undec}
$\existsprob$ is undecidable.
\end{restatable}
\begin{proof}
We prove this theorem by contradiction.  Suppose, if possible, there exists a halting Turing machine $M$ that takes as inputs a vocabulary $\VV$, a $\VV$-formula $\xi(\Z)$ and a computable $\VV$-structure $\MM$, and decides if there exists a computable $\MM$-interpretation of Skolem functions for all existential variables in $\xi(\Z)$. We show below that we can use $M$ to effectively decide if an arbitrary Turing machine, say $\TM{i}$, halts on the empty tape. 

Consider $\VV = \{Q, a\}$, where $Q$ 
is a binary predicate symbol and $a$ is a constant symbol.  For each $i \in \mathbb{N}$, define $\MM_i$ to be a $\VV$-structure such that $\U^{\MM_i} = \mathbb{N}$, $a^{\MM_i} = i$ and $Q^{\MM_i}(u, v) = \ttrue$ for $u, v \in \mathbb{N}$ iff the Turing machine $\TM{u}$ halts on the empty tape within $v$ steps.  It is easy to see that each $\MM_j$ is a computable $\VV$-structure.  We also define the $\VV$-sentence
$\xi \equiv \exists s \forall u \exists x\, \big( \big(Q(a, s) \wedge (x = u)\big)   \vee \big(\neg Q(a, u) \wedge Q(u, x)\big) \big)$.
Skolemizing this formula gives
$\xi^\star \equiv \forall u\, \big( \big( Q(a, c_s) \wedge (F_x(u) = u)\big) \vee \big(\neg Q(a,u) \wedge Q(u, F_x(u))\big)\big),$
where 
$c_s$ is a Skolem constant for $s$, and $F_x$ is a Skolem function for $x$.
From the guarantee of Skolemization, the following must hold:
\begin{itemize}
\item $\exists s \forall u \exists x\, \big( \big(Q(a, s) \wedge (x = u)\big)   \vee \big(\neg Q(a, u) \wedge Q(u, x)\big) \big)$ $~\Leftrightarrow~ $ $\forall u \exists x\, \big( \big(Q(a, c_s) \wedge (x = u)\big)   \vee \big(\neg Q(a, u) \wedge Q(u, x)\big) \big)$
\item $\forall u \big( \exists x\, \big( \big(Q(a, c_s) \wedge (x = u)\big)   \vee \big(\neg Q(a, u) \wedge Q(u, x)\big) \big)$ $\Leftrightarrow$ $\big( \big(Q(a, c_s) \wedge (F_x(u) = u)\big)   \vee \big(\neg Q(a, u) \wedge Q(u, F_x(u))\big) \big) \big)$
\end{itemize}
We now consider two cases.
\begin{itemize}
    \item Suppose $\TM{i}$ halts on the empty tape after $p \in \mathbb{N}$ steps. Then $\MM_i \models Q(a, p)$.  In this case, by choosing the $c_s^{\MM_i} = p$ and by choosing $F_x^{\MM_i}(u) = u$ for all $u \in \mathbb{N}$, both the above guarantees of Skolemization are easily seen to hold.  Clearly, the Skolem functions have computable interpretations in this case.
    \item If $\TM{i}$ doesn't halt on the empty tape, then $\MM_i \models \forall u\, \neg Q(a, u)$.    In this case,
    we choose an arbitary value, say $0$, for $s$.  However, for the guarantee of Skolemization to hold, we must have the following: for every $u \in \mathbb{N}$, if $\exists x Q(u, x)$ holds (i.e. $\TM{u}$ halts on the empty tape), then $Q(u, F_x^{\MM_i}(u))$ must also hold  (i.e. $\TM{u}$ must also halt in $F_x^{\MM_i}(u)$ steps).  Clearly, such an interpretation $F^{\MM_i}_x$ is not computable, as otherwise it can be used to decide the halting problem.
\end{itemize}
The above reasoning shows that there exist computable $\MM_i$-interpretations of all Skolem functions of existential variables in $\xi$ iff $\TM{i}$ halts on empty tape. Thus, if we feed the instance $(\VV, \MM_i, \xi)$ as input to the supposed Turing machine $M$ that decides $\existsprob$, we can decide if $\TM{i}$ halts on the empty tape, for every $i \in \mathbb{N}$. This gives a decision procedure for the halting problem on the empty tape -- an impossibility!

\end{proof}

It is easy to see that the proof of Theorem~\ref{thm:existsprob_undec} can be repeated with $\xi \equiv \forall u \exists s \exists x\, \big( \big(Q(a, s) \wedge (x = u)\big) \vee \big(\neg Q(a, u) \wedge Q(u, x)\big) \big)$ as well.  This gives the following interesting result.
\begin{restatable}{theorem}{quant_prefix}\label{thm:quant_prefix}
If the vocabulary contains a binary predicate and a constant, $\existsprob$ is undecidable for the quantifier prefix classes $\exists\forall\exists$ and $\forall\exists\exists$. However it is decidable for the class $\exists^+\forall^*$ 
\end{restatable}
The second part of the above Theorem follows from an easy generalization of the proof of Lemma~\ref{lem:existsprob_const}.  This leaves only the case of $\forall\exists$ quantifier prefix, for which the decidability of $\existsprob$ remains open. 
We consider the case of the vocabulary having only monadic predicates later in Theorem~\ref{thm:lowenheim}. 

The above negative results motivate us to consider special cases of $\existsprob$ and $\synthprob$, where either the $\VV$-formula $\xi(\Z)$ or the $\VV$-structure $\MM$ is fixed.

%% file: undec.tex
The proof of Theorem~\ref{thm:existsprob_undec} is quite damning:  even if we allow the possibility of a potentially different algorithm, say $A_{\VV,\xi}$, for deciding $\existsprob$ for each combination of $\VV$ and $\xi$, we cannot hope to have an algorithm $A_{\VV,\xi}$ for every $(\VV,\xi)$ pair.  This is because in the proof of Theorem~\ref{thm:existsprob_undec}, we had indeed kept the vocabulary and formula fixed. This leaves only a few questions to be investigated if we fix the vocabulary and formula.
If we consider $\VV$ and $\xi$ as fixed, the $\VV$-structure $\MM$ is the only input to our problems of interest.   The following theorem shows that $\synthprob$ cannot be answered positively in this case even under fairly strong conditions.

Recall from Lemma~\ref{lem:existsprob_const} that $\existsprob$ has a positive answer if all Skolem functions are Skolem constants. Hence, by choosing $\xi$ to be a $\VV$-sentence with only existential quantifiers, we are guaranteed that all problem instances are positive instances of $\existsprob$.
\begin{restatable}{theorem}{uniform}
\label{thm:uniform}
There exists a vocabulary $\VV$, a $\VV$-sentence $\xi$ and a family of $\VV$-structures $\Ff = \{\MM_i \mid i \in \mathbb{N}\}$, such that $(\VV, \MM_i, \xi)$ is a positive instance of $\existsprob$ for all $i \in \mathbb{N}$, yet there is no uniform representation of $\MM_i$-interpretations of the Skolem constants.  Additionally, the {\synthprob} problem has a negative answer for the class of problem instances $\{(\VV, \xi, \MM_i) \mid i \in \mathbb{N}\}$.
\end{restatable}
\begin{proof}
Let $\VV = \{P\}$, where $P$ is a unary predicate.  We consider the $\VV$-formula $\xi \equiv \exists x\, P(x)$.  Each $\VV$-structure $\MM_i$ is defined as follows: $U^{\MM_i} = \mathbb{N}$ and $P^{\MM_i}(u) = \ttrue$ iff the Turing machine $\TM{i}$ halts on the empty tape within $u$ steps.  Note that $P^{\MM_i}(u)$ is effectively computable for every $i, v \in \mathbb{N}$.  Therefore, every $\MM_i$ is a computable structure.  The Skolemized version of $\xi$ is $\xi^\star \equiv P(c_x)$, where $c_x$ is the Skolem constant for $x$.

We know from Lemma~\ref{lem:existsprob_const} that there exists a computable $\MM_i$-interpretation of $c_x$ for every $\MM_i$.  If possible, let there be a uniform representation of these $\MM_i$-interpretations.  In other words, there exists a halting Turing machine, say $M$, with access to an oracle $\mathcal{O}_P$ that provides the interpretation of $P$, such that $M$ computes the value of $c_x$ for $\MM_i$, for all $i \in \mathbb{N}$.  We show below that this implies the existence of a Turing machine $M^h$ that decides the halting problem on the empty tape.

Turing machine $M^h$ takes as input $i \in \mathbb{N}$ and mimics the steps of $M$ until an invocation of $\mathcal{O}_P$ happens, say with argument $u\in \mathbb{N}$.  $M^h$ then mimics the steps of $\TM{i}$ for upto $u$ steps, and if $\TM{i}$ halts (resp. does not halt) by then, $M^h$ resumes mimicking the steps of $M$ assuming that the oracle invocation returned $\ttrue$ (resp. $\ffalse$).  Since $M$ is a halting Turing machine, the computation of $M^h$ effectively computes the $\MM_i$-interpretation of $c_x$, say $j$. The machine $M^h$ then mimics the steps of $\TM{i}$ for upto $j$ steps and checks whether it halts by then.  If it does, $M^h$ outputs Yes, otherwise it outputs No.

From the guarantee of Skolemization, we know that $\MM_i \models \big( \exists x\, P(x) ~\Leftrightarrow~ P(j)\big)$, where $j$ is as computed above.  Therefore, indeed if $P^{\MM_i}(j)$ is $\ffalse$, we must have $\MM_i \models \exists x\, P(x)$.  This shows the correctness of the output computed by $M^h$.

Since the halting problem on the empty tape is undecidable, this proves that Turing machine $M$ cannot exist.  In other words, there is no uniform representation of the $\MM_i$-intepretations of the (only) Skolem constant.

To show that the {\synthprob} problem also has a negative answer in this case, suppose there exists a halting Turing machine $M'$ that accepts $\MM_i$ as input and generates a constant $j \in \mathbb{N}$ such that $j$ serves as an $\MM_i$-interpretation of $c_x$ in $\xi^\star$.  Using a similar reasoning as above, we can construct a halting Turing machine, say $\widehat{M^h}$, that mimics the steps of $M'$ to compute $j$ and then mimics the steps of $\TM{i}$ for up to $j$ steps to check if $\TM{i}$ halts on the empty tape in at most $j$ steps.  Clearly, $\widehat{M^h}$ would serve as a decision procedure for the halting problem on the empty tape --  an impossibility! Hence the Turing machine $M'$ cannot exist.  It follows that $\synthprob$ has a negative answer for the class of problem instances $\{(\VV, \MM_i, \xi) \mid i \in \mathbb{N}\}$.
\end{proof}

 It is interesting to ask now if there is a characterization of $\VV$-formulas, such that for each $\VV$-formula satisfying this characterization, the $\existsprob$ and $\synthprob$ problems have positive answers for all $\VV$-structures. The proof of Theorem~\ref{thm:existsprob_neg} tells us that we must disallow binary predicates and  $\forall\exists$ blocks in the quantifier prefix, which severely restricts the vocabulary and formulas.
What happens if we allow a relational vocabulary with only monadic predicates (Lowenheim class with equality)~\cite{BGG97}? 
\begin{restatable}{theorem}{lowenheim}\label{thm:lowenheim}
Let the vocabulary $\VV$ contain only monadic predicates and equality. Then $\existsprob$ has a positive answer, but not so for $\synthprob$. \end{restatable}

\begin{proof}
With $k$ monadic predicates, the universe can be partitioned into $2^k$ equivalence classes based on predicate valuations. By an argument (based on Ehrenfeucht-Fraisse games) similar to that used to prove small-model property of Lowenheim class (see~\cite{BGG97}), if a prenex formula $\xi$ has quantifier rank $r$, the range of each Skolem function can be restricted to $\le r.2^k$ elements. Using an argument similar to that in proof of Lemma~\ref{lem:existsprob_const}, there exists a TM that enumerates the required set, say $S$, of $\leq r.2^k$ elements. Since elements of an equivalence class can only be distinguished using $=$, for each Skolem function of arity $p$, we must search for its correct interpretation over all $S^p \rightarrow S$ mappings. Since there are finitely many such mappings, we can enumerate the TMs computing these mappings, and one of them must effectively serve as the correct interpretation for the Skolem function under consideration. Since $\xi$ has finitely many existential variables, it follows that there exists computable interpretations of all Skolem functions in $\xi$. 

To see why $\synthprob$ has a negative answer in general even with one monadic predicate $P$ and one existential quantifier, consider $V=\{P\}$, $\xi \equiv \exists x P(x)$, and $M_i$ having universe $\mathbb{N}$ and $P^{M_i}(x)=$true iff $TM_i$ halts on empty tape within $x$ steps. If there exists an algorithm to synthesize computable $M_i$-interpretations of the Skolem constant for $x$ in $\xi$, we can use it to decide if $TM_i$ halts on empty tape -- an impossibility! Thus, we must disallow even monadic predicates if we want to characterize V-formulas that admit positive answer to $\synthprob$ for all $\VV$-structures. 
\end{proof}

%% file: fixed-structure.tex
\medskip 

\noindent{\bf Fixing the structure.}
We now fix the structure $\MM$ (and vocabulary $\VV$) and take the formula $\xi$ as the only input of our problems of interest. Since the structure $\MM$ is fixed, we use the notation $\univ$ for $U^{\MM}$ henceforth. Theorem~\ref{thm:existsprob_neg} already shows that even when the structure is fixed, the {\existsprob} problem has a negative instance.  However, the $\VV$-structure used in that proof may appear hand-crafted. This leads us to ask if there is a "natural" vocabulary $\VV$ and $\VV$-structure $\MM$, such that {\existsprob} has a negative instance when considering $\VV$-formulas. It turns out that this is indeed the case, and we show it by appealing to the classical Matiyasevich-Robinson-Davis-Putnam (MRDP) theorem~\cite{MRDP}.

\begin{restatable}{proposition}{MRDP}
\label{prop:MRDP}
Skolem functions for the first order theory of natural numbers over the vocabulary $\{\times, +,0,1\}$ 
do not admit computable interpretations.
\end{restatable}
\begin{proof}
  The proof in this case appeals to the Matiyasevich-Robinson-Davis-Putnam (MRDP) theorem~\cite{MRDP} that equates Diophantine sets with recursively
enumerable sets. Formally, it states: \emph{A set of natural
numbers is Diophantine if and only if
it is recursively enumerable}. Recall that a set $\mathcal{S}$ of
natural numbers is Diophantine if there exists a polynomial $P(x,
y_1, \ldots y_k)$ with integer coefficients such that the Diophantine
equation $P(x, y_1, \ldots y_k) = 0$ has a solution in the unknowns
$y_1, \ldots y_k$ iff $x \in \mathcal{S}$.  Recall also that a set
$\mathcal{S}$ is recursively enumerable if there exists a (potentially
non-halting) Turing machine that outputs every element of
$\mathcal{S}$ in some order, and only those elements.  Now consider
the set $S_{\mathit{halt}}$ of natural number encodings of all Turing
machines that halt on the empty tape.  It is a well-known result in
computability theory~\cite{TOCbook} that $S_{\mathit{halt}}$ is
recursively enumerable, although there is no Turing machine that takes
a natural number $x$ as input and halts and correctly reports whether
$x \in S_{\mathit{halt}}$.  By the MRDP theorem, recursive
enumerability of $S_{\mathit{halt}}$ implies the existence of a
polynomial $P_{\mathit{halt}}(x, y_1, \ldots y_k)$ such that
$x \in S_{\mathit{halt}}$
iff $\exists y_1, \ldots y_k \in \mathbb{N}^k\,
P_{\mathit{halt}}(x, y_1, \ldots y_k) = 0$.  We now consider the first
order sentence $\xi_{\mathit{halt}} = \forall x \exists
y_1 \ldots \exists y_k\, P_{\mathit{halt}}(x, y_1, \ldots y_k) =
0$. Note that since $P_{\mathit{halt}}(x, y_1, \ldots y_k)$ is a
polynomial, it can be written as a term in the first order theory of
natural numbers with vocabulary $\{\times, +, 0, 1\}$.  Furthermore, on
Skolemizing, we get $\xi^\star_{\mathit{halt}} = \forall x\,
P_{\mathit{halt}}(x, f_1(x), \ldots f_k(x))$, where $f_1, \ldots f_k$
are unary Skolem functions.  Suppose, if possible, there exist Turing
machines $M^1_{\mathit{halt}}$ through $M^k_{\mathit{halt}}$ that take
$x \in \mathbb{N}$ as input, and always halt and compute the values of
$f_1(x)$ through $f_k(x)$ respectively. Given $x \in \mathbb{N}$, we
can then use $M^1_{\mathit{halt}}$ through $M^k_{\mathit{halt}}$ to
compute the values of $f_1(x), \ldots f_k(x)$, and determine if
$P_{\mathit{halt}}(x, f_1(x), \ldots f_k(x)) = 0$.  If so, we know
that $x \in S_{\mathit{halt}}$; otherwise $x \not\in
S_{\mathit{halt}}$. This gives an algorithm (or halting Turing machine) to
determine if any natural number $x \in S_{\mathit{halt}}$ -- an
impossibility!  Hence, there cannot exist Turing machines
$M^1_{\mathit{halt}}$ through $M^k_{\mathit{halt}}$ that compute the
Skolem functions $f_1(x)$ through $f_k(x)$.
      \end{proof}      
Finally, in the setting of a fixed $\VV$-structure $\MM$, even if $\existsprob$ is answered in the positive for all $\VV$-formulas in a class $\Xi$, the $\synthprob$ problem may have a negative answer for the class of problem instances $\{(\VV, \MM, \xi) \mid \xi \in \Xi\}$.

\begin{restatable}{theorem}{synthneg}
\label{thm:synth_neg}
There exists a vocabulary $\VV$, a $\VV$-structure $\MM$ and a class of $\VV$-sentences $\Xi = \{\xi_i \mid i \in \mathbb{N}\}$ s.t., $(\VV, \MM, \xi)$ is a positive instance of $\existsprob$ for all $\xi_i \in \Xi$, yet $\synthprob$ has a negative answer for the class of instances $\{(\VV, \MM, \xi_i) \mid i \in \mathbb{N}\}$.
\end{restatable}
\begin{proof}
Consider the vocabulary $\VV = \{Q, f, a\}$, where $Q$ is a binary predicate, $f$ is a unary function and $a$ is a constant symbol.  Let $\MM$ be the $\VV$-structure with $\Univ = \mathbb{N}$, $Q^{\MM}(u,v) = \ttrue$ iff Turing machine $\TM{u}$ halts on the empty tape within $v$ steps, $f^{\MM}(u)$ is the successor of $u$ and $a^{\MM} = 0$, for all $u, v \in \mathbb{N}$. For each $i \in \mathbb{N}$, we also define $\xi_i \equiv \exists v\, Q(f^{(i)}(a), v)$, where $f^{(i)}$ denotes $i$ nested applications of $f$. 

Note that since every $\xi_i$ has only existential variables, by Lemma~\ref{lem:existsprob_const}, $(\VV, \MM, \xi_i)$ is a positive instance of $\existsprob$ for all $i \in \mathbb{N}$.

Let $\xi_i^\star \equiv Q(f^{(i)}(a), c_v)$ be the Skolemized version of $\xi_i$, where $c_v$ is a Skolem constant for $v$.
Suppose $\synthprob$ has a positive answer for the class of problem instances $\{(\VV, \MM, \xi_i) \mid i \in \mathbb{N}\}$. Then, there exists a halting Turing machine, say $M$, that takes as input $\VV, \MM$ and $\xi_i$ and outputs the encoding of a Turing machine, say $N_i$, that halts on the empty tape and outputs a natural number, say $j \in \mathbb{N}$, that serves as an $\MM$-interpretation of $c_v$. Since $f^{(i)}(a)$ interprets to $i$ in $\MM$, then by the guarantee of Skolemization, it is easy to show that we can now check if $\TM{i}$ halts on the empty tape by simply simulating a run of $\TM{i}$ for upto $j$ steps. %
We now construct a Turing machine, say $\widehat{M}$ that accepts as input $i \in \mathbb{N}$, constructs the problem instance $(\VV, \MM, \xi_i)$ and then mimics the run of the above Turing machine $M$ on $(\VV, \MM, \xi_i)$.  Since $M$ always halts, the mimicking of $M$ by $\widehat{M}$ must eventually stop. The machine $\widehat{M}$ then uses the output, say $j \in \mathbb{N}$, computed by $M$ to mimic the run of $\TM{i}$ for at most $j$ steps.  If during this process, $\widehat{M}$ detects that $\TM{i}$, it outputs Yes; else it outputs No.  It is easy to see from the construction above that for every $i \in \mathbb{N}$, the Turing machine halts and reports Yes iff $\TM{i}$ halts on the empty tape.  Since the halting problem on the empty tape is undecidable, this shows that Turing machine $M$ cannot exist.

In other words, $\synthprob$ has a negative answer.
\end{proof}

\section{Necessary and sufficient condition for synthesizing Skolem functions}
\label{sec:charac}
Given these strong negative results is there hope for proving existence and synthesizability of  computable interpretations for Skolem functions. 
Indeed, there do exist many natural theories where computable interpretations of Skolem functions exist and can indeed be synthesized, e.g., Boolean case, Presburger arithmetic etc. 
So, what determines when a $\VV$-theory admits effective synthesis of computable interpretations of Skolem functions for all $\VV$-formulas? 
Our first positive result is a surprising characterization of a \emph{necessary and sufficient} condition for algorithmic synthesis of computable interpretations of Skolem functions.

\begin{restatable}{theorem}{charac}
\label{thm:charac}
Let $\MM$ be a computable $\VV$-structure for vocabulary $\VV$. The $\synthprob$ problem for $\VV$-formulas, i.e. for problem instances $\{(\VV, \MM, \xi) \mid \xi$ is a $\VV$-formula$\}$, has a positive answer iff $\eldiag{\MM}$ is decidable.
\end{restatable}
\begin{proof}
$(\Longleftarrow)$ 
Let $\xi(\Z)$ be a $\VV$-formula with free variables $\Z$, where $\xi(\Z)\equiv \forall \X_1 \exists \Y_1 \ldots \forall \X_n \exists\Y_n\\ \xi_n (\Z,\X_1,\Y_1,$$\ldots,$ $\X_n,\Y_n)$, where $\X_1,\ldots,\X_n$ are $n$ sequences of universally quantified variables, $\Y_1,\ldots \Y_n$ are sequences of existentially quantified variables, $\Z$ is a sequence of free variables and $\xi_n$ is quantifier-free. 
We will show that there is an algorithm, that for every $i\in\{1,\ldots, n\}$, takes as input a $(|\Z|+|\X_1|+\ldots +|\X_i|)$ tuple of values from the universe $\univ$, say, $\mu\in \univ^{|\Z|}, \sigma_1\in \univ^{|\X_1|},\ldots , \sigma_i\in \univ^{|\X_i|}$ and halts after computing a $(|\Y_1|+\ldots +|\Y_i|)$-dimensional vector of values, $\bF_1(\mu,\sigma_1)\in \univ^{|Y_1|},\ldots \bF_i(\mu, \sigma_1,\ldots \sigma_i)\in \univ^{|\Y_i|}$ where for each $1\leq j\leq i$, $\bF_j$ is a $|\Y_j|$-dimensional vector of Skolem functions, each of arity $|\Z|+|\X_1|+\ldots +|\X_j|$.

The proof is by induction on $i$.  For $i=1$, let $\xi(\Z) \equiv \forall \X_1 \exists \Y_1\, \xi_1(\Z, \X_1,\Y_1)$, where $\xi_1$ has one less number of  quantifier alternations than $\xi$. 
On Skolemizing, we get $\xi^\star(\Z) \equiv \forall \X_1\, \xi_1(\Z,\X_1,\bF_1(\Z,\X_1))$, where $\bF_1$ is a $|\Y_1|$-dimensional vector of Skolem functions each of arity $|\Z|+|\X_1|$.  We now design a Turing machine (or algorithm) $M_1$ that
takes any $|\Z|+|\X_1|$-tuple of elements from $\univ$, say
$(\mu,\sigma_1)$, as input and halts after computing $\bF_1(\mu,\sigma_1)$:
\begin{itemize}
\item[(a)] It first determines if $\exists \Y_1\, \xi_1(\mu,\sigma_1, \Y_1)$ holds, using the decision procedure for $\eldiag{\MM}$.
\item[(b)] If the answer to the above question is ``Yes'', the machine $M_1$
recursively enumerates $|\Y_1|$-tuples of elements of $\univ$, and for each
tuple $\nu$ thus enumerated, it checks if $\xi_1(\mu,\sigma_1,\nu)$ evaluates to true. Again the  decidability of $\eldiag{\MM}$ ensures that this check can also be effectively done.  The machine $M_1$ outputs the
first (in recursive enumeration order) element of $\univ^{|\Y_1|}$, for which $\xi_1(\mu,\sigma_1,\nu)$ is true as $\bF_1(\mu,\sigma_1)$, and halts.
\item[(c)] If the answer is ``No'', i.e. there is
no $\nu \in \univ^{|\Y_1|}$ s.t. $\xi_1(\mu,\sigma_1,\nu)$ is true, $M_1$ outputs the first (in recursive enumeration order) tuple of
$\univ^{|\Y_1|}$ as $\bF_1(\mu,\sigma_1)$, and halts.
\end{itemize}
It is easy to verify that the vector of functions $\bF_1$ computed by $M_1$ satisfies
$\forall \X_1\,\big(\exists \Y_1\, \xi_1(\Z, \X_1,\Y_1) \Leftrightarrow \xi_1(\Z,\X_1, \bF_1(\Z,\X_1))\big)$ for every valuation of the free variables $\Z$ in
$\univ^{|\Z|}$, i.e., we have a (correct)  $\MM$-interpretation of Skolem function $\bF_1$. This completes the base case for $i=1$.

For the general case of $i \ge 1$, we write $\xi(\Z)$ as $\forall \X_1 \exists \Y_1\ldots$ $\forall \X_i\exists \Y_i \xi_i(\Z,\X_1,\Y_1,\ldots \X_i,\Y_i)$, where  $\xi_i$ is a formula with $i$ less $\forall^*\exists^*$ blocks than $\xi$. By induction hypothesis, we know that there exists a Turing machine $M_i$ that takes as input any values for free variables $\Z$ and universally quantified variables $\X_1,\ldots\X_i$ and outputs values for  $\Y_1,\ldots\Y_i$ so that they correspond to outputs of (correct) $\MM$-interpretations of vectors of Skolem functions $\bF_1,\ldots \bF_i$. 

We need to show the existence of a computable interpretation of the vector of Skolem functions $\bF_{i+1}$ for $\Y_{i+1}$. Thus, we are given a $(|\Z|+|\X_1|+\ldots +|\X_i|+|\X_{i+1}|)$-tuple of values from $\univ$, and we need to show how to define a Turing machine $M_{i+1}$ that takes this vector as input and halts after computing values of  vectors of Skolem functions $\bF_1(\mu,\sigma_1), \ldots \bF_{i+1}(\mu, \sigma_1,\ldots \sigma_i,\sigma_{i+1})$.
Let $(\mu,\sigma_1,\ldots,\sigma_i,\sigma_{i+1})$ be the given set of input values. The Turing machine $M_{i+1}$ first simulates $M_i$ on input $(\mu,\sigma_1,\ldots,\sigma_i)$. This returns $i$ vectors of values $\nu_1=\bF_1(\mu,\sigma_1)\in \univ^{|Y_1|}, \ldots \nu_i=\bF_i(\mu, \sigma_1,\ldots \sigma_i)\in \univ^{|Y_i|}$ such that each of $\bF_1,\ldots, \bF_i$ is a Skolem function vector. Plugging in all these values in $\xi_i$ gives a sentence $\hat{\xi}_i=\xi_i[\Z\mapsto \mu, \X_1\mapsto \sigma_1, \Y_1\mapsto \nu_1,\ldots,\X_i\mapsto \sigma_i,\Y_i\mapsto \nu_i]$. Observe that $\hat{\xi}_i\equiv \forall \X_{i+1}\exists \Y_{i+1}\hat{\xi}_{i+1}(\X_{i+1},\Y_{i+1})$ for some formula $\hat{\xi}_{i+1}(\X_{i+1},\Y_{i+1})$.
We can then apply the same argument as in the base case above and conclude.

    Note that the values of $\Z, \X_1, \ldots \X_n$ used above were arbitrary tuples from $\univ^{|\Z|}$, $\univ^{|\X_1|}, \ldots \univ^{|\X_n|}$. Hence lifting the notation introduced in  sentence~(\ref{fact1}) of Section~\ref{sec:prelim} to talk about vector of variables $\Y_{i+1}$ instead of single variables $y_{i,j}$, we conclude that the vector of functions $\bF_{i+1}$ computed by Turing machine $M_{i+1}$ satisfies $\forall \Z \forall \X_1 \ldots \forall \X_{i+1}\,\\ \big(\exists \Y_{i+1} \xi^\star_{\Y_{i+1}} ~\Leftrightarrow~ \xi^\star_{\Y_{i+1}}[\Y_{i+1}\mapsto \bF_{i+1}(\Z, \X_1, \ldots \X_{i+1})]\big)$. Thus, $\bF_{i+1}$ gives a (correct) $\MM$-interpretation of a vector of Skolem functions for $\Y_{i+1}$, which completes the proof.

($\implies$) In the other direction, we will show that if there exists a halting Turing machine, say $M$, that synthesizes computable $\MM$-interpretations of Skolem functions for all existential variables in all $\VV$-formulas, then $\eldiag{\MM}$ must be decidable. We assume that there are at least two elements in $\univ$; let's call them $d_1$, $d_2$. 
To show that $\eldiag{\MM}$ is decidable, we need to show a decision procedure for the $\expvoc{\VV}{\MM}$-theory of $\modelconst{\MM}$.   Consider any $\expvoc{\VV}{\MM}$-sentence $\varphi$. This sentence may have finitely many constants that are not in $\VV$. Let us say  $c_1,\ldots c_k$ are these constants, for some non-negative integer $k$. We introduce $k$ fresh variables $y_1,\ldots y_k$ and define $\varphi'$ to be the formula obtained by taking $\varphi$ and replacing each occurrence of the constant $c_i$ by variable $y_i$ respectively. Note that $\varphi'$ is a $\VV$-formula.
We introduce 3 more fresh variables $x,z_1,z_2$ and consider the sentence $\psi\equiv \forall y_1\ldots \forall y_k\forall z_1\forall z_2 \exists x\big (((x=z_1)\wedge \varphi')\vee ((x=z_2)\wedge \neg \varphi')\big)$.
We can easily rewrite this formula in prenex normal form, but doing so still leaves $x$ as the leftmost existentially quantified variable. We now feed this prenex normal form of $\psi$ as the input to Turing machine $M$ (that synthesizes computable interpretations of Skolem functions for all existential variables in all $\VV$-formulas).  Let the computable interpretation of the Skolem function for $x$, as output by $M$, be the Turing machine $M_x$.  Therefore, $M_x$ takes as inputs values of $y_1,\ldots y_k, z_1,z_2$ (as these are the only universally quantified variables to the left of $x$ in the quantifier prefix) and it always halts after computing a value for $x$. Now we run the Turing machine $M_x$ with the inputs: $c_i$ as the value for $y_i$ for $i \in \{1,\ldots k\}$, $d_1$ as the value for $z_1$ and $d_2$ as the value for $z_2$.  Let the value computed by $M_x$ with these inputs be $t$. We then check if $t=d_1$. If yes, we conclude that $\modelconst{\MM} \models \varphi$, else $\modelconst{\MM} \not\models \varphi$. Thus, we have a decision procedure for $\expvoc{\VV}{\MM}$-theory of $\modelconst{\MM}$, i.e., $\eldiag{\MM}$ is decidable.
\end{proof}
The construction in the first part of the above proof shows that there exists a Turing machine that runs in time polynomial in the length of $\xi$ and in the length of a decision procedure for $\eldiag{\MM}$, and generates a computable interpretation (i.e. code for a halting Turing machine) that computes $\MM$-interpretations of all Skolem functions in $\xi$.  Further, since a positive answer to $\synthprob$ implies a positive answer to $\existsprob$, Theorem~\ref{thm:charac} also gives a sufficient condition for $\existsprob$ to have a positive answer.

\section{Applications and complexity}
\label{sec:applications}
We now look at some consequences of the above characterization. We first ask if we can algorithmically synthesize computable interpretations of Skolem functions in some well-known theories in first-order logic. We consider the following theories of interest.
\begin{enumerate}
    \item Presburger arithmetic:  first-order theory of natural numbers with addition and order; $\VV=\{+,<,0,1\}$ and $\univ = \mathbb{N}$.
    \item Linear rational arithmetic: first-order theory of rationals with addition and order; $\VV=\{+,<,0,1\}$ and $\univ = \mathbb{Q}$.
    \item Theory of reals: first-order theory of real algebraic numbers with addition, multiplication and order;
    $\VV=\{\times,+,<,0,1\}$ and $\univ =$ set of real algebraic numbers.
    \item Theory of countable dense linear order without endpoints: This is unique up to isomorphism, and is canonically modeled by the rational numbers; $\VV = \{<\}$ and $\univ = \mathbb{Q}$.
    \item Peano arithmetic: first-order theory of natural numbers with addition, multiplication and order; $\univ=\{\times, +,<,0,1\}$ and $\univ = \mathbb{N}$.
\end{enumerate}

We start with a lemma.
\begin{restatable}{lemma}{transfer}
\label{lem:transfer}
Let $\MM$ be a computable $\VV$-structure with universe $\univ$. Suppose for every element $e\in \univ$, there exists an effectively computable uni-variate $\VV$-formula $\alpha_e(x)$ such that $\alpha_e(x)$ is $\ttrue$ iff $x=e$. Then $\Theory{\MM}$  is decidable iff $\eldiag{\MM}$ is decidable.
\end{restatable}
\begin{proof}
Recall from Section~\ref{sec:modeltheory} that $\eldiag{\MM}$ is the $\expvoc{\VV}{\MM}$-theory of $\modelconst{\MM}$. Therefore, one direction of the proof is trivial: if $\eldiag{\MM}$ is decidable, then so is $\Theory{\MM}$. To show the other direction, consider an arbitrary $\expvoc{\VV}{\MM}$-sentence $\varphi$. Recall from Section~\ref{sec:modeltheory} that $\expvoc{\VV}{\MM}$ has a separate constant symbol, say $c_e$, for every $e \in \univ$. For every such constant $c_e$ that occurs in $\varphi$ but is not in $\VV$, we introduce a fresh variable $v_e$. Given that the formula is finite, we only have finitely many such variables.  Let $S$ denote set of all such variables. We then define $\varphi'$ to be $\varphi$ in which every occurrence of $c_e$, where $e \in \univ$, is replaced by the corresponding variable $v_e$.  Now we consider 
$\xi\equiv \exists_{v_e \in S}~ v_e\, \big(\varphi'\wedge  \bigwedge_{v_e\in S}  \alpha_e (v_e)\big)$.

Since $\xi$ is a $\VV$-sentence, we can appeal to the decision procedure for $\Theory{\MM}$ to check if $\MM \models \xi$. It is easy to see that if the answer is yes, then $\modelconst{\MM} \models \varphi$, i.e. $\varphi \in \eldiag{\MM}$. Otherwise, $\modelconst{\MM} \not\models \varphi$, i.e. $\varphi \not\in \eldiag{ \MM}$. Thus we have obtained a decision procedure for $\eldiag{\MM}$ using one for $\Theory{\MM}$.
\end{proof}
One may wonder if decidability of $\Theory{\MM}$ automatically implies decidability of $\eldiag{\MM}$. However, this is not true in general, emphasizing the need for Lemma~\ref{lem:transfer}.

\begin{proposition}
\label{prop:equality}
There exists a vocabulary $\VV$ and $\VV$-structure $\MM$ such that $\Theory{\MM}$ is decidable but $\Theory{\modelconst{\MM}}$ is not. 
\end{proposition}
\begin{proof}
We will use as before that Turing machines can be identified with elements of $\mathbb{N}$. Consider the vocabulary $\VV = {Q, R, f}$, where $Q, R$ are unary predicates and $f$ is a unary function.  Consider the $\VV$-structure $\MM$ is defined as follows:
 The universe is $\{u\#1v \mid  u, v \in \{0,1\}^+\}\cup \{u \mid u\in \{0,1\}^+\}$

We define the interpretations of $Q$ and $R$ as follows:
\begin{itemize}
\item $Q(x) = true$ iff $x$ has a $\#1$ in it.
\item $R(x) = true$ iff $Q(x)$ is true and if $x = u\#1v$, then the Turing machine represented by $u$ halts in at most $v$ steps when started on the empty tape.
\end{itemize}

Define the interpretation of $f$ as follows:
\begin{itemize}
\item  If $Q(x)$ is true, then $f(x) = u$, where $x = u\#1v$
\item If $Q(x)$ is false, then $f(x)=x$
\end{itemize}

Note that there are no infinite chains of application of $f$.  Specifically, $f(f(x)) = f(x)$ for all $x$. Clearly, the interpretations of $Q(x)$, $R(x)$ and $f(x)$ are all computable.  In other words, there are halting Turning machines that given $x$, can compute $Q(x)$, $R(x)$ and $f(x)$ for all $x$ in the universe of $\MM$.

Now we claim that $\Theory{\modelconst{\MM}}$ is undecidable.  To see this consider  
for a Turing machine encoded by the binary number $c \in \{0,1\}^+$, the formula $\exists x (Q(x) \wedge R(x) \wedge (f(x) = c))$. This formula  is satisfiable in $\modelconst{\MM}$ iff the Turing machine represented by $c$ halts on the empty tape.  A decision procedure for $\Theory{\modelconst{\MM}}$ therefore decides the halting problem.

However, $\Theory{\MM}$ is decidable. This can be shown on the lines of the proof of decidability of Lowenheim's theory with equality~(see for e.g.,\cite{BGG97}). There is a small difference however, since in the proof of the small model property of Lowenheim's theory with equality, the interpretations of the unary functions may need to be changed. However, a very similar idea works here because of the special structure of the interpretation of $f$, i.e. $f(f(x)) = f(x)$ for all $x$. So, given a sentence $\varphi$ with quantifier rank (i.e., number of quantifiers) $r$ in $\Theory{\MM}$, choose the substructure $\MM'$ of $\MM$ induced by the following elements of its universe: $r$ distinct $u\#1v$ such that TM $u$ halts on the empty tape in at most $v$ steps, $r$ distinct $u\#1v$ s.t. TM $u$ doesn't halt on the empty tape in $v$ steps, and the $2r$ distinct $u$'s from the above choices of $u\#1v$, and $r$ additional distinct $u$'s (that don't appear in the above $u\#1v$'s).  For an Ehrenfeucht-Fraisse (EF) game of $r$ rounds, clearly the duplicator now has a winning strategy.  Therefore, $\MM \models \varphi$ iff $\MM' \models \varphi$.  However, $\MM'$ being a finite structure, we have an effective procedure to decide if $\MM' \models \varphi$.  This gives a decision procedure to decide in $\MM \models \varphi$ or equivalently if $\Theory{\MM} \models \varphi$.

Therefore, we can indeed have $\Theory{\MM}$ decidable, but $\Theory{\modelconst{\MM}}$ undecidable, even when all predicates and functions in $\modelconst{\MM}$ are computable.
\end{proof}

From Lemma~\ref{lem:transfer} and Theorem~\ref{thm:charac} we now have,

\begin{restatable}{corollary}{appl}
\label{cor:appl}
For the following theories, both \existsprob\ and \synthprob\ have  positive answers, and we can effectively synthesize computable $\MM$-interpretations for Skolem functions for a $\VV$-formula:
    1. Presburger arithmetic;
    2. Linear rational arithmetic (LRA);
    3. Theory of real algebraic numbers;
    4. Theory of dense linear orders without endpoints.
\end{restatable}

\begin{proof}
  In cases (1-3), $\alpha_e(x)$ as required for Lemma~\ref{lem:transfer} can be easily defined: (i) for Presburger arithmetic, for each integer $k$, $\alpha_k(x)$ is $x=0+1 +1\cdots+ 1$ ($k$ times), (ii) for LRA, for any rational $p/q$, $\alpha_{p/q}$ is $q.x = p$, and (iii) for real algebraic numbers, we know that for every algebraic number $a$, there exists a uni-variate polynomial, say $p_a(x)$, with integral coefficients and a pair of rational bounds that define an interval containing exactly one root of $p_a(x)$, i.e. $a$. Furthermore, there are decision procedures for Presburger arithmetic~\cite{Pre29}, LRA~\cite{decbook} and theory of real algebraic numbers~\cite{Tarski51}. So by  Lemma~\ref{lem:transfer}, $\eldiag{\MM}$ is decidable in all these cases. The result now follows from Theorem~\ref{thm:charac}.
  
  For case 4, the result follows by observing that (i) the theory of dense linear order without endpoints is isomorphic to $(\mathbb{Q},<)$, and (ii) the elementary diagram of $(\mathbb{Q},<)$ is contained in that of LRA, which we have shown is decidable. 
\end{proof}

For the theory of natural numbers with addition, multiplication and order, we have seen in Proposition~\ref{prop:MRDP} that  \existsprob\ has a negative answer, which of course implies that \synthprob\ cannot have a positive answer. Using Lemma~\ref{lem:transfer} and Theorem~\ref{thm:charac} we get a direct proof for the latter fact. To see this note that the premise of Lemma~\ref{lem:transfer} holds for this theory as for Presburger arithmetic. Hence, $\eldiag{\MM}$ is decidable iff $\Theory{\MM}$ is decidable. But we know from the MRDP theorem~\cite{MRDP} that the latter is indeed undecidable. Thus, from Theorem~\ref{thm:charac}, we obtain that \synthprob\ has negative instances in this theory. 

We remark that the above discussion can also be seen as an alternate proof of the fact that the elementary diagram is undecidable, since Theorem~\ref{thm:charac} is a characterization.

\medskip 

%% file: complexity.tex
\noindent{\bf Complexity bounds on  $\MM$-interpretations.}
When it applies, the proof of Theorem~\ref{thm:charac} gives us a construction of a Turing machine $M$ that takes a formula $\xi$ as input and outputs a computable $\MM$-interpretation (i.e. code for another Turing machine, say $M'$) of Skolem functions for all existential variables in $\xi$.
What bounds can we give on the worst case running time of $M'$ (Problem 4.c in Section~\ref{sec:problem})? 
We start with a lower bound that follows from the second part of the proof of Theorem~\ref{thm:charac}.
\begin{restatable}{theorem}{lowerbound}Let $\MM$ be a computable $\VV$-structure with a decidable $\eldiag{\MM}$. The worst case running time of any computable $\MM$-interpretation of Skolem functions for a $\VV$-formula is at least as much as that of a decision procedure for $\eldiag{\MM}$.
\end{restatable}

This shows for instance that for Presburger arithmetic, there exists formulas for which any computable $\MM$-interpretation of  Skolem functions will take at least (alternating) double exponential time~\cite{berman80,Haase18} (also see Section~\ref{app:presburger}).
Next, for upper bounds, the computable $\MM$-interpretation of Skolem functions, as detailed in the proof of Theorem~\ref{thm:charac}, relies on enumeration. Hence, it does not help in giving complexity upper bounds. However, if a theory admits effective constraint solving (see Sec.~\ref{sec:prelim} for a definition), then we can do better. 
\begin{restatable}{theorem}{complexity}
\label{thm:complexity}
Let $\MM$ be a $\VV$-structure such that $\eldiag{\MM}$ is decidable. Suppose  $\eldiag{\MM}$ admits effective constraint solving with worst-case time complexity $T(n)$ and the computed result being of size at most $S(n)$. Then we can synthesize  $\MM$-interpretations of Skolem functions for $\VV$-formulas of size $n$, such that the running time and output size of the $\MM$-interpretations are bounded by recursive functions of $T(n)$ and $S(n)$.
\end{restatable}

\begin{proof}

Let $\xi(\Z)\equiv \forall \X_1 \exists \Y_1 \ldots \X_k\Y_k \xi_k(\Z, \X_1,\Y_1, \ldots \Y_k)$. Given a $|\Z|+|\X_1|+\cdots |\X_k|$-dimensional vector of values $(\mu,\sigma_1,\ldots \sigma_k)$ as input, we substitute $\mu$ for $\Z$ and $\sigma_1$ for $\X_1$ in $\xi_k$ and consider $\xi'(\Y_1)\equiv \forall \X_2 \exists \Y_2\ldots \X_k\Y_k \xi_k(\mu,\sigma_1,\Y_1,\X_2, \ldots \Y_k)$. In $\xi'$,  $\Y_1$ are free variables and we call the constraint solving algorithm which returns a vector of values $\nu_1$ for $\Y_1$ such that $\exists \Y_1\xi'(\Y_1) \equiv \subst{\xi'}{\Y_1}{\nu}$. Let $\xi_1=\xi'[\Y_1\mapsto \nu]$. We output this $\nu$ as the value of the Skolem function vector for $\Y_1$ and continue with $\xi_1$. By recursively repeating the above steps for $\xi_1$, we can obtain values for all existential variables. This effectively gives a computable $\MM$-interpretation of all Skolem functions.

To analyze the complexity, let $t_i$ (resp. $s_i$) denote the time taken and size of output when processing the $i^{th}$ $\forall^*\exists^*$ block in the above steps. We begin wtih $s_0=n$, where $n$ is the size of the formula $\xi$ and $t_0=T(s_0)=T(n)$. Now for every $i\geq 0$, we have (because of substitutions) the following recurrence relation between sizes of the formula at step $i$ and $i+1$: $s_{i+1}=s_i+S(s_i)$ and the time for processing the $i^{th}$ $\forall^*\exists^*$ block is given by $t_i=T(s_i)$. Simplifying, we get that the total time is $time(n) =\sum_{i=1}^k t_i $ and total size of the Skolem functions is $size(n)= s_k(n)$. 

\end{proof}
As an example, if $S(n)$ is linear, i.e., $S(n)\leq C.n$ for a constant $C > 0$, then we get  $ time(n) \leq k.T(k.(\max\{C,1\})^{k}.n)$ and $size(n)\leq k.(\max\{C,1\})^k.n$. Finally, one way to obtain an algorithm for effective constraint solving is by using effective quantifier elimination repeatedly and then using quantifier-free constraint solving. Thus, we could further bound the complexity as functions of the complexity for effective quantifier elimination and that of quantifier-free constraint solving. This can be  applied, for example, for LRA, theory of reals etc. Significantly, there are first-order theories that do not admit effective quantifier elimination but admit effective constraint solving, e.g., theory of evaluated trees~\cite{DK06}. In such cases, we can still use our approach to synthesize Skolem functions. 

\subsection{A brief note on Presburger arithmetic}
\label{app:presburger}
From Corollary~\ref{cor:appl}, we know that we can effectively synthesize Skolem functions for Presburger arithmetic. Presburger arithmetic as such does not admit quantifier elimination (e.g., $\exists x (y=x+x)$ has no quantifier free equivalent), although adding divisibility predicates allows us to get back quantifier elimination. If we follow this approach, then after applying quantifier elimination, we would need not quantifier-free constraint solving, but constraint solving with uni-variate existential formulas representing  divisibility predicates.  One can also use an automata-based approach for effective constraint solving, using ideas from automata-based decidability of Presburger arithmetic~\cite{wolper95,habermehl10}.
This suggests a novel automata-theoretic approach for synthesizing computable interpretations of Skolem functions for Presburger formulas. We believe this deserves further exploration and could lead to practically efficient implementations of Skolem functions.

\section{Expressing Skolem functions as terms}
\label{sec:skolemterm}
Whenever Skolem functions are computable, one can further ask: \emph{Can Skolem functions be represented as terms?}  Notice that in the Boolean setting, the notions of terms, functions and formulas are often conflated (as noted by Jiang~\cite{Jian} as well). Note that there are theories without any terms, for which Skolem functions can still be synthesized as halting Turing machines. For instance, the theory of (countable) dense linear order without endpoints does not admit any terms. Yet, from Corollary~\ref{cor:appl}, we know that we can effectively synthesize computable Skolem functions for this theory. In fact, we can show a stronger result, viz, even when the theory admits terms, we may not be able to interpret a Skolem function as a term. To see this, consider the Presburger formula:
$\forall y \forall z \exists x (((x=y)\vee (x=z))\wedge ((x\geq y) \wedge (x\geq z)))$.
The unique Skolem function for $x$ is $\max(y,z)$, which can be written as an imperative program as:
"{\bfseries input(y,z); if $\mathtt{y\geq z}$ then return $y$ else return $z$}".
This is a uniform representation (see Sec.~\ref{sec:prelim}) of a computable $\MM$-interpretation of the Skolem function for $x$. However, this function cannot be written as a term in Presburger arithmetic. Indeed, any term of $y$, $z$ that uses only $+$, $0$, $1$ must be linear, while max is a non-linear function.  Thus, we have
\begin{proposition}
There exist first order theories for which Skolem functions can be effectively computed, but they cannot be expressed as terms.
\end{proposition}
As described in~\cite{Jian}, if Skolem functions in a first order theory can be represented using a \emph{finite set of conditional terms} (like in the case of $\max(y,z)$ above), the theory admits effective quantifier elimination.  However, we already have first order theories, e.g. the theory of evaluated trees, that don't admit quantifier elimination, but admit effective synthesis of computable interpretations of Skolem functions. In such cases, Skolem functions can't be represented as a finite set of conditional terms either.

%% file: conclusion.tex
The study of algorithmic computation of Skolem functions is highly nuanced. We explored what it means for Skolem functions for first order logic to be computable and synthesizable. Defining computable interpretations of Skolem functions as Turing machines, we showed that they may not always exist and checking if they exist is undecidable in general. However, when we fix a computable structure, we gave a precise characterization of when they exist and show several applications for specific theories. While we have made some preliminary progress regarding complexity issues, the question of synthesizing succinct interpretations is still open as is the question of when Skolem functions can be represented as terms in the logic. We hope that the theoretical framework set up here will lead to research towards implementable synthesis of Skolem functions for first order logic.